\documentclass[11pt]{article}

\usepackage{amsmath}
\usepackage{amssymb}
\usepackage{graphicx}
\usepackage{float}
\usepackage[hdivide={20mm,,20mm}, vdivide={30mm,,30mm}, nohead]{geometry}
\usepackage[dvips]{color}
\usepackage{colordvi}
\usepackage{url}
\usepackage{wrapfig}
\usepackage[normalem]{ulem}  
\usepackage{comment}

\allowdisplaybreaks[1]
\newcommand{\bt}{{\beta}}
\newcommand{\ga}{{\gamma}}

\newcommand{\te}{{\theta}}

\newcommand{\bsx}{{\boldsymbol{x}}}

\newcommand{\bfR}{{\mathbf{R}}}

\newcommand{\cL}{{\mathcal{L}}}

\newcommand{\nn}{{\nonumber}}
\newcommand{\ol}{\overline}
\newcommand{\pd}{{\partial}}

\newcommand{\ran}{{\rangle}}

\newcommand{\wh}{\widehat}

\def\det{\mathop{\rm det}\nolimits}

\def\tr{\mathop{\rm tr}\nolimits}
\def\diag{\mathop{\rm diag}\nolimits}
\def\Re{\mathop{\rm Re}\nolimits}

\def\bbra{{\langle\kern-2.5pt\langle}}
\def\kket{{\rangle\kern-2.5pt\rangle}}
\def\Bbra{{\Big\langle\kern-3.5pt\Big\langle}}
\def\Kket{{\Big\rangle\kern-3.5pt\Big\rangle}}

\def\ran{\rangle}
\def\eq#1{(\ref{#1})}

\begin{document}
\baselineskip=18pt

\thispagestyle{empty}
\addtocounter{page}{-1}
{}
\vskip-5cm
\begin{flushright}
TIFR/TH/15-23
\end{flushright}
\vspace*{0.7cm} \centerline{\Large \bf 
Thermodynamics of QCD from Sakai-Sugimoto Model}
\vspace*{1 cm} 
\centerline{\bf 
Hiroshi~Isono$^1$, Gautam~Mandal$^2$ and Takeshi~Morita$^3$}
\vspace*{0.5cm}
\centerline{\rm $^1$\it Department of Physics,}
\centerline{\it National Tsing Hua University,} 
\centerline{\it Hsinchu 30013, \rm TAIWAN}
\centerline{\tt email: hiroshi.isono81@gmail.com}
\vspace*{0.3cm}
\centerline{\rm $^2$\it Department of Theoretical Physics,}
\centerline{\it Tata Institute of Fundamental Research,} 
\centerline{\it Mumbai 400 005, \rm INDIA}
\centerline{\tt email: mandal@theory.tifr.res.in}
\vspace*{0.3cm}
\centerline{\rm $^3$\rm \it Department of Physics}
\centerline{\it Shizuoka University,} 
\centerline{\it 836 Ohya, Suruga-ku, Shizuoka 422-8529,  \rm JAPAN}
\centerline{\tt email: morita.takeshi@shizuoka.ac.jp}

\vspace*{1cm}
\centerline{\bf Abstract}
\vspace*{0.5cm} 
\enlargethispage{1000pt}

Till date, the only consistent description of the deconfinement phase
of the Sakai-Sugimoto model appears to be provided by the analysis of
\cite{Mandal:2011ws}. The current version of the analysis, however,
has a subtlety regarding the monodromy of quarks around the Euclidean time
circle. In this note, we revisit and resolve this issue by considering
the effect of an imaginary baryon chemical potential on quark
monodromies. With this ingredient, the proposal of
\cite{Mandal:2011ws} for investigating finite temperature QCD using
holography is firmly established. Additionally, our technique allows a
holographic computation of the free energy as a function of the
imaginary chemical potential in the deconfinement phase; we show that
our result agrees with the corresponding formula obtained from
perturbative QCD, namely the Roberge-Weiss potential.

\newpage

\section{Introduction}

The Witten-Sakai-Sugimoto model \cite{Witten:1998zw, Sakai:2004cn} has
been of profound significance in investigating QCD using holography
\cite{Maldacena:1997re}. Witten's model for a holographic dual of pure
Yang-Mills theory \cite{Witten:1998zw} involved essentially $N_c$ D4
branes on $\mathbf{R}^3 \times$ a Scherk-Schwarz circle. Sakai and Sugimoto
\cite{Sakai:2004cn} added external fundamental quarks to this model by
adding $N_f$ D8 and $\overline{{\rm D8}}$-branes as probes and
demonstrated that this model explains the low energy dynamics of QCD
such as chiral symmetry breaking and the meson spectrum. By now, a
large body of literature has grown (see, {\it e.g}, \cite{Kim:2012ey,
  Rebhan:2014rxa} for recent developments) which successfully explains
various qualitative aspects of QCD at strong coupling using this
model.

It is, thus, natural to apply this model to systems at finite
temperature. Such a study, indeed, was initiated by Aharony et al
\cite{Aharony:2006da}, who, following \cite{Aharony:1999ti}, employed
the black D4-brane geometry as the gravity dual to the deconfinement
phase in QCD and explained chiral symmetry restoration at high
temperature.  This was followed by many authors who investigated this
model to explore various outstanding problems in thermal QCD such as
the phase structure including the finite baryon chemical potential,
which is difficult to compute in lattice gauge theory due to the
well-known sign problem \cite{Kim:2006gp, Horigome:2006xu, Sin:2007ze,
  Yamada:2007ys, Bergman:2007wp}.

Ref. \cite{Aharony:2006da}, however, also pointed out that the
correspondence between the black D4-brane geometry and the
deconfinement phase in QCD is not without problems, since there must
be at least one phase boundary between these two phases, which,
therefore, are not smoothly connected. This issue was further explored
by two of the present authors (MM) \cite{Mandal:2011ws}, who showed
that the black D4-brane geometry cannot describe four dimensional QCD
due to issues related to the large-$N_c$ volume independence
\cite{Eguchi:1982nm, Gocksch:1982en} (as applied to the Scherk-Schwarz
circle).  Instead, \cite{Mandal:2011ws} proposed that the
deconfinement phase in QCD is represented by a rather different
geometry, involving a localized solitonic D3-brane. It was shown that
the gravity solution and deconfined QCD belong to the same phase; in
addition, \cite{Mandal:2011ws} showed that the new geometry too offers
a mechanism of chiral symmetry restoration ($\chi$SR).\footnote{Like
  in \cite{Aharony:2006da}, here too $\chi$SR is preceded by the
  deconfinement transition, although unlike in \cite{Aharony:2006da},
  the latter is a Gregory-Laflamme transition.}

There have been many follow-up works on finite temperature QCD from
holography which support the MM proposal. In particular, it is argued
in \cite{Rebhan:2014rxa} that the estimate of the deconfinement
transition temperature obtained from MM fits phenomenology better than
the one obtained from the Witten-Sakai-Sugimoto model
\cite{Witten:1998zw, Sakai:2004cn}. The papers \cite{Dasgupta:2014ffa,
  Chen:2012me} argue that an alternative description of the high
temperature phase of QCD is possible in a more elaborate setup using
D4, NS5 and D6, $\overline{\rm D6}$ branes, although within the
Sakai-Sugimoto setup the high temperature phase indeed follows MM.
Besides, Ref.\,\cite{Azuma:2014cfa} demonstrated the similarity between the confinement/deconfinement transition in the gauge theory and the phase transition in the MM model, and a related connection can be seen even analytically in supersymmetric Yang-Mills theory \cite{Morita:2014ypa}.
More recently, Ref.\,\cite{Hanada:2015gsa} showed that the MM model
qualitatively reproduces the instanton density in lattice QCD while
the black D4-brane geometry does not.

The MM construction of the deconfinement phase, however, has an
unsolved question regarding the monodromy of quarks around the
temporal circle; we will discuss this below in Section
\ref{sec-MM}. The purpose of this article is to resolve
this issue and thus firmly establish the MM method
\cite{Mandal:2011ws} for investigating finite temperature QCD via
holography.  We achieve this resolution by considering, as already
hinted at in MM, an imaginary baryon chemical potential
\cite{Roberge:1986mm} (see Sections \ref{impot} and \ref{impotss}).
We find that this leads to the correct monodromy of the quarks
suitable for describing thermal physics.  As an additional application
of our technique, we compute in Section \ref{sec-theta-dep} the
dependence of the free energy on the imaginary chemical potential; we
find that the result is consistent with the Roberge-Weiss potential
\cite{Roberge:1986mm}\footnote{See, e.g., \cite{Kashiwa:2015tna} for some 
recent progress on QCD with complex chemical potential.} 
which is obtained from perturbative QCD at high temperature.

\section{Deconfinement phase in Sakai-Sugimoto model and 
the issue of quark monodromy}
\label{sec-MM}

To obtain ${\rm SU}(N_c)$ QCD with $N_f$ flavor quarks, Sakai and
Sugimoto \cite{Sakai:2004cn} considered the following brane
configuration
\begin{eqnarray}
\begin{array}{lcccccccccc}
& (t) & 1 & 2 & 3 & (4) & 5 & 6 & 7 & 8 & 9 \\
N_c~ \text{D4-brane} & - & - & - & - & - &&&&& \\
N_f~ {\rm D8/\overline{D8}}\text{-brane}
& - & - & - & - &  & - & - & - &- & - 
\end{array}
\label{config-IIA}
\end{eqnarray}
Here the parentheses denote directions compactified on $S^1$. To
represent finite temperature, spacetime is regarded as Euclidean; $t$
parameterizes the Euclidean time circle with period $\beta=1/T$.
$x^4$ is the Scherk-Schwarz circle,  with antiperiodic boundary
condition for fermions, which is crucial to break supersymmetry and to
obtain the (non-supersymmetric) four dimensional QCD at low energy
\cite{Witten:1998zw}.  We define the period of this circle 
to be $L_4$.

We will consider below $N_f \ll N_c$ so that we can treat the
D8-branes as probes.  In an appropriate large $N_c$ limit {\em \`{a}
  la} Maldacena \cite{Maldacena:1997re, Itzhaki:1998dd} we can replace
the D4 branes by a classical geometry, and regard the D8 branes as
probes coupled to such a geometry. At low temperatures, the geometry,
which in fact is the dual gravity description to compactified 5
dimensional SYM theory \cite{Itzhaki:1998dd}, is given by
\begin{align}
ds^2 =& \alpha' \left[\frac{u^{3/2}}{\sqrt{ \lambda_5/4\pi }}
\left( dt^2  + \sum_{i=1}^{3} 
dx_i^2+f_4(u) dx_4^2 \right)+ \frac{\sqrt{ \lambda_5/4\pi }}{u^{3/2}}\left(  
\frac{du^2}{f_4(u)} 
+ u^2 d\Omega_{4}^2 \right)  \right], \nonumber \\
& f_4(u)=1-\left( \frac{u_0}{u}\right)^3 , \quad e^{\phi}=\frac{\lambda_5}{(2\pi)^2N_c}
\left(\frac{u^{3/2}}{\sqrt{\lambda_5/4\pi}}  \right)^{1/2}.
 \label{metric-SD4}
\end{align}
There is also a non-trivial value of the five form potential which we
do not write explicitly.  Here $\lambda_{5} := (2\pi)^{2} g_s l_s N_c$
is the 't Hooft coupling on the D$4$-brane world-volume.  $g_s$ and
$l_s= \sqrt{\alpha'}$ are the string coupling and the string length,
respectively.  We will also use the dimensionless coupling
$\lambda_{YM} := 2\lambda_5/L_4$. The identifications $t \equiv t+
\beta, x_4 \equiv x_4 + L_4$ are implicit.

Since the $x_4$-cycle shrinks to zero at $u=u_0$, in order to avoid
possible conical singularities we must choose $L_4$ as follows 
\begin{align}
\frac{L_4}{2\pi}= \frac{\sqrt{\lambda_{5}/4\pi}}{3} u_0^{-1/2} .
\label{u0-beta}
\end{align} 
With this choice, the contractible $x_4$-cycle, together with the
radial direction $u$, forms a so-called `cigar' geometry which is
topologically a disc. 

Note that the above gravity solution is reliable in a regime
$\lambda_{YM} \gg 1$ in which the stringy corrections are suppressed.
On the other hand in the regime $\lambda_{YM} \ll 1$, the system flows
to IR and the massive modes such as adjoint fermions (and scalars),
together with KK modes about the $x^4$-circle are decoupled and we
obtain four dimensional QCD \cite{Sakai:2004cn}.\footnote{The theory
  we obtain is actually four dimensional pure Yang Mills theory
  \cite{Witten:1998zw} with external quarks in the fundamental
  representation coming from the D4-D8($\overline{\rm D8}$) open
  strings.} It turns out, however, that although there is no direct
overlap of validity between four dimensional QCD at weak coupling
($\lambda_{YM} \ll 1$) and gravity at strong coupling ($\lambda_{YM}
\gg 1$), it has been a highly rewarding enterprise to understand QCD
dynamics by extrapolation from the gravity analysis (see, {\it e.g},
\cite{Kim:2012ey,Rebhan:2014rxa} for recent developments).  We should
note that such an extrapolation is analogous to the use of the strong
coupling expansion in lattice gauge theory. An important consideration
for the success of the latter expansion is that there should not be a
phase boundary between the strong coupling theory and the continuum
theory which is defined at zero lattice coupling
\cite{Polyakov:1987ez}. The point of the MM proposal
\cite{Mandal:2011ws} was to ensure that such phase boundaries were not
encountered between gravity and four-dimensional QCD either at low or
at high temperature. 

To investigate high temperature regime using the gravity dual in the
sense mentioned above, \cite{Mandal:2011ws} pointed out that it was
imperative to impose the periodic boundary condition along the
temporal circle for the fermions.\footnote{\label{ftnt:bc}The usual
  antiperiodic boundary condition leads to the black D4 brane at high
  temperature, which is separated from four dimensional deconfined QCD
  by a phase boundary, whereas the use of the periodic boundary
  condition leads to a different gravity solution, as we will recall
  presently, which does not involve such a phase boundary. The
  consequence of this choice for the description of high temperature
  QCD is detailed at the end of this section, and is the main issue
  addressed in this paper.}  As we increase the temperature, strings
which wind the temporal circle become lighter, and the type IIA
gravity analysis is not reliable when the temperature reaches around
$T \sim \sqrt{\lambda_{YM}}/ L_4$. This leads us to a type IIB
supergravity description obtained by performing a T-duality on the
temporal circle. \footnote{If we had taken the periodicity of the
  fermions along the temporal circle to be anti-periodic, the system
  would be mapped to type 0B rather than to type IIB.} Then the
analysis is reliable if $T \gg \sqrt{\lambda_{YM}}/ L_4$ \footnote{To
  be more precise, the mass of the winding string depends on its
  position, and the lightest one is located at $u=u_0$ in the IIA
  flame.  However after the T-duality, since the radius is flipped,
  the heaviest one is from $u=u_0$ in the IIB flame. Thus, if $T \gg
  \sqrt{\lambda_{YM}}/ L_4$, while the IIA supergravity analysis
  around $u \sim u_0$ is not reliable, the IIB supergravity is,
  although the validity of the latter away from this region is not
  ensured.  However, since it is only the region $u\sim u_0$ which is
  involved in exploring the confinement/deconfinement transition, the
  use of IIB supergravity in $T \gg \sqrt{\lambda_{YM}}/ L_4$ is
  justified.}.  This T-duality maps the brane configuration of the
Sakai-Sugimoto model (\ref{config-IIA}) to
\begin{eqnarray}
\begin{array}{lcccccccccc}
& (t') & 1 & 2 & 3 & (4) & 5 & 6 & 7 & 8 & 9 \\
N_c~ \text{D3-brane} &  & - & - & - & - &&&&& \\
N_f~ {\rm D7/\overline{D7}}\text{-brane}
&  & - & - & - &  & - & - & - &- & - 
\end{array}
\label{config-IIB}
\end{eqnarray}
Here $t'$ is the dual temporal circle, and its periodicity is 
$ \beta' = (2\pi)^2/\beta=(2\pi)^2T $.

The IIB supergravity analysis shows that around
\begin{align}
T_{GL} \sim \frac{\lambda_4}{ L_4}
\label{GL-t}
\end{align}
the T-dual of the gravity solution (\ref{metric-SD4}) becomes unstable
due to the Gregory-Laflamme instability \cite{Gregory:1994bj}, and a
Gregory-Laflamme transition occurs.  The stable solution at higher
temperatures is the localized solitonic D3-brane geometry.  Generally
the analysis of the localized geometry in a compact space is
difficult; however, an approximate geometry of the localized solitonic 
D3-brane geometry, can be described by the metric \cite{Mandal:2011ws}
\begin{align}
ds^2 &= \alpha' \left[H^{-1/2}
\left( \sum_{i=1}^{3}  
dx_i^2 + (1+2\Phi)dx_4^2  \right)
 + H^{1/2} (1-\frac{1}{2}\Phi)\left(  du^2 +dt'^2+ u^2 d\Omega_{4}^2  \right)  \right]  , \nonumber \\ 
&
H= \sum_n \frac{2 \lambda_5/\beta }{(u^{2}+(t'-t'_0-n \beta')^{2})^2}, \quad 
 e^{\phi}=\frac{\lambda_5}{2\pi N_c \beta}, \nonumber \\
& \Phi=- \frac{u^4_H}{2} \sum_n \left( \frac{1}{u^2+(t'-t'_0-n \beta')^2} \right)^2, \quad 
 u_H= \sqrt{2 \lambda_5 T}\frac{\pi}{2 L_4}.  
\label{metric-LD3-large-u}
\end{align}
The approximation used here is valid in a region far from the
localized D3-branes and for $T \gg u_H$.  Note that this solution
describes D3-branes localized around $t'=t'_0$ and it breaks the
translation symmetry along the $t'$-circle spontaneously.  We fix
$t'_0=0$ in the following discussions.
\cite{Mandal:2011ws} argued that it is this localized D3-brane
geometry that corresponds to the deconfinement phase in QCD, and the
Gregory-Laflamme transition leading to this localized solution
corresponds to the confinement/deconfinement transition.\footnote{The
  spontaneous symmetry breaking of the translation symmetry
  corresponds to the breaking of the center symmetry in the
  deconfinement phase.}

To study quarks in the deconfinement phase, we need to study the probe
D7-branes (\ref{config-IIB}) on the background geometry
(\ref{metric-LD3-large-u}).  However, we appear to encounter a problem
here. As we explained above, we have imposed a periodic spin structure
(around the temporal circle) in our supergravity analysis. Naively,
this would seem to
suggest\footnote{\label{ftnt:spin-structure}Strictly speaking, the
  periodic spin structure in gravity refers to gravitinos and other
  bulk fermions.  To connect it to the spin structure of the gauge
  theory, we need to consider the AdS/CFT dictionary which relates
  gauge invariant boundary fermion operators to bulk fermions. The
  {\it naive} assumption referred to here is that the spin structure
  of the gauge theory is simply the same as that of supergravity. As
  we will see below, in our model they differ by an imaginary baryon
  chemical potential.} periodic BC on all QCD fermions, while the
standard description of thermodynamics requires the BC to be
antiperiodic. The thermodynamics of the pure Yang-Mills sector is not
affected by this problem, since at weak coupling, the adjoint fermions
on the D3-branes (D4-branes in the IIA description) are anyway
decoupled from the theory, hence the choice of their temporal boundary
condition is immaterial \cite{Mandal:2011ws}. However, since the
fundamental quarks of QCD in the Sakai-Sugimoto model are not
decoupled in this limit, a periodic temporal BC for them would appear
to be at variance with standard rules of QCD thermodynamics.
The resolution of the above problem is a central issue of this paper.

\section{Imaginary baryon chemical potential and quark monodromy}\label{impot}

In this section, we consider the issue of monodromy of fermions in QFT
in somewhat more general terms. Let us suppose that the quarks $\psi$
in four dimensional $SU(N_c)$ QCD satisfy the following boundary
condition around the temporal circle\footnote{The temporal circle is Euclidean and is labelled by $t$. Note that $\theta=\pi$
  corresponds to the periodic boundary condition and $\theta=0$ to the
  anti-periodic one.},
\begin{align}
\psi(t+\beta)=-e^{-i \theta} \psi(t). 
\label{BC}
\end{align} 
As shown in Appendix \ref{app-chemical}, a path
integral with the above boundary condition can be interpreted as 
a thermal partition function with an imaginary baryon chemical
potential ($\mu= i \theta$)
\begin{align}
Z(\beta,\theta) := \tr \left( e^{-\beta \hat{H} + i \theta \hat{N} } \right),  
\label{thermal-partition-QCD}
\end{align}
Here $\hat{N}$ is the baryon number operator. In Appendix
\ref{app-chemical} and what follows, we show how such a chemical
potential arises from in the presence of a thermal Wilson line or,
equivalently, from the temporal component a constant $U(1)$ external
gauge potential. We will indicate in the next section, how such a
gauge potential is automatically present in our problem (it is
determined by the location of the D7 brane on the temporal circle). We
will find that 
\\ (a) Choosing the above parameter (location of the D7
brane) judiciously, we can obtain a thermal partition function
starting from a path integral over periodic fermions (Section
\ref{impotss}). This resolves the issue of apparent conflict between
periodic fermion BC and thermodynamics.  
\\ (b) Keeping this parameter
arbitrary, the holographic setup allows us to determine the dependence
of the free energy as a function of the imaginary chemical potential
(Section \ref{sec-theta-dep}).  The computation agrees with the
corresponding computation from perturbative QCD--- a result called the
the Roberge-Weiss potential.

\section{The imaginary chemical potential in the Sakai-Sugimoto model}\label{impotss}

We will now show how an imaginary baryon chemical potential such as in
\eq{thermal-partition-QCD} presents itself in the Sakai-Sugimoto model
in our context (see \cite{Rafferty:2011hd} for a related
discussion).\footnote{Note that the
  imaginary chemical potential does not affect the periodicity of the
  color adjoint fermions and they remain periodic.}
 Let us first recall that the standard (real) chemical potential in Euclidean SU($N_c$) QCD 
is equivalent to a constant pure-imaginary temporal
component of the Euclideanised diagonal flavor gauge field.
Hence an imaginary chemical potential is related to the real-valued Euclideanised flavor gauge field (see \eqref{diracpart} for a precise relation between the two). 
In the Sakai-Sugimoto model, the flavor gauge field corresponds to a gauge field on D8/$\overline{\rm D8}$-branes. 
Combining this observation with the fact that fermions on the string theory side are periodic in our setup, we find that {\it a thermal partition function with an imaginary
  chemical potential $\theta$ can be regarded as a periodic functional integral
 with an external flavor gauge potential $A^f_t$ through}
\begin{align}
\bt A_{t}^{f} = \theta + \pi.
\label{chemical-gauge}
\end{align}
Note that the factor $\pi$ comes from the difference of the periodicities. 
We give a detailed derivation of \eqref{chemical-gauge} in Appendix \ref{derive}.

We discussed the imaginary chemical potential above (and in Appendix
\ref{app-chemical}) directly in terms of quarks in the D4-D8 open
strings. In the dual supergravity theory, the quarks are not present
directly, but baryons are. It is easy to demonstrate how baryons pick
up nontrivial monodromies around the thermal circle in the presence of
a constant diagonal flavor gauge field. A baryon in the gravity dual
is represented by a `baryon vertex' (a D4-brane which wraps the $S^4$
of the geometry (\ref{metric-SD4})) \cite{Sakai:2004cn}. The D4-brane
couples to the diagonal flavor gauge field (considered as external)
through a Chern-Simons term as
\begin{align}
\exp
\left( i \int A \wedge F_4 \right) \sim \exp \left( i N_c \int_0^\beta
dt \, A_t^{f} \right).
\label{baryon}
\end{align}
The gravity functional integral in the presence of this D4-brane
defines the baryon wavefunction. As one takes $t \to t+\beta$, the
functional integral, and hence the baryon wavefunction, picks up an
extra phase $e^{i N_c \beta A^f_t}$ due to the presence of the
flavor gauge field.

Thus, to introduce an arbitrary imaginary chemical potential, we need
to simply turn on an appropriate constant, diagonal, flavor gauge
field $A^f_t$, according to \eq{chemical-gauge}. To proceed
further, we need to do a bit more work. \eq{chemical-gauge} is written
in terms of type IIA variables.  As explained in Section \ref{sec-MM},
to describe the high temperature limit in our holographic framework
(corresponding to deconfined QCD) we need to T-dualize on the temporal
circle. In that case, $2\pi A_{t}^{f}$ is mapped to the position
$t'_f$~\footnote{In the holographic setup $t'_f$ corresponds to the
  point on the $t'$-circle reached by the D7 brane at the boundary,
  {\it i.e.} $t'_f=$$\lim_{_{u\to \infty}} t'(u)$.} of the
D7/$\overline{\rm D7}$-branes on the dual temporal circle. The type
IIB version of \eq{chemical-gauge} is, then\footnote{The length of the
  temporal circle in the IIB is denoted by $\beta'$ which equals
  $4\pi^2/\beta$.}
\begin{align}
t'_f= \frac{2\pi}{\beta}(\pi+\theta) = \frac{\beta'}{2\pi}(\pi+\theta).
\label{position-theta}
\end{align}
In particular, we can satisfy the above condition with (this was
indeed the choice made in \cite{Mandal:2011ws})
\begin{align}
t'_f=  \beta'/2 ~~~~ (\Rightarrow ~~ \theta=0).
\label{antipodal}
\end{align}
Recall that by convention we have put the D3 branes at
$t'=0$. Equation \eq{antipodal} implies that if the D7/$\overline{\rm
  D7}$-branes are placed antipodally from the D3 branes on the
temporal circle of the IIB theory (equivalent to a choice $\theta=0$),
the path integral with periodic fermions {\it does reduce to a thermal
  partition function}, thus resolving the apparent puzzle raised in
Section \ref{sec-MM}.

One might wonder whether we can place the D7 branes at will at any
point on the temporal circle. In case the D7 branes are infinitely
heavy, the position of the D7's can be regarded as an external
parameter. However, in the next section, we will consider the
potential energy of the D7 branes, placed at a position $t'_f$ at the
boundary ($u \to \infty$).\footnote{This can be regarded as the
  potential energy between the D7 and D3 branes in the open string
  description.} We will find that the choice $\theta=0$ mentioned
above is indeed a minimum of the potential.\footnote{See 
\eq{pot-theta-holo}. Other minima, corresponding to non-zero
values of the integer $k$, correspond to the presence of $k$
D3-brane pairs as explained in footnote \ref{ftnt:schwinger}.}

\section{\label{sec-theta-dep}Free energy as a function of
imaginary chemical potential}

\begin{figure}[H]
\begin{center}
\includegraphics[scale=.40]{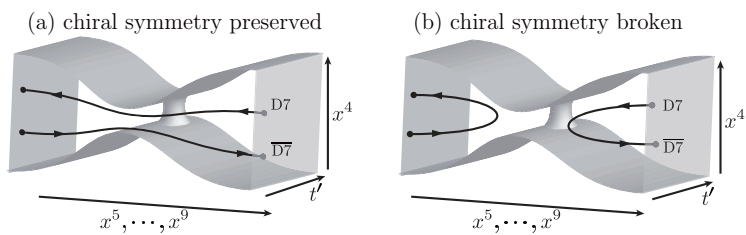}
\caption{D7/$\overline{{\rm D7}}$-branes in the localized solitonic
  D3-brane background (\ref{metric-LD3-large-u}). There are two
  possible configurations of the D7/$\overline{{\rm D7}}$-branes: (a)
  the D7 and $\overline{{\rm D7}}$-branes extend parallel to each
  other, which implies that the chiral symmetry is preserved, (b) the
  D7 and $\overline{{\rm D7}}$-branes are connected, which implies that
  the chiral symmetry is broken.}
\label{fig-D7}
\end{center}
\end{figure}

In this section we will calculate the free energy of the D3-D7 system
(primarily as a function of the location $t'_f$ of the D7 branes on
the temporal circle).  As mentioned before, we will use the dual
gravity description (\ref{metric-LD3-large-u}) for the D3 branes and
consider the D7 branes as probes coupled to this geometry through a
DBI action.  In the deconfinement geometry (\ref{metric-LD3-large-u}),
there are two possible stable configurations of the D7-branes which
correspond to the chiral symmetry being preserved/broken
\cite{Mandal:2011ws} (see Fig.~\ref{fig-D7}).

\noindent Let us consider the chiral-symmetric configuration for
simplicity; it is straightforward to generalize our considerations to
the broken symmetry configuration.  We assume that the branes are not
curved in $x_4$ space.  Then the induced metric on the D7-branes is
given by
\begin{align}
ds^2_{\rm D7} =& \alpha' \left[H^{-1/2}
\left( \sum_{i=1}^{3} 
dx_i^2 \right)+H^{1/2}\left( \left(1+ \left(\frac{d t'(u)}{du}\right)^2 \right) du^2  + u^2 d\Omega_{4}^2 \right)   \right], \nonumber
\\ 
H=& \sum_n \frac{2 \lambda_5/\beta }{(u^{2}+(t'(u)-n \beta')^{2})^2} \nonumber \\
=&\frac{ \lambda_5}{4\pi }  \left[ \frac{1}{ u^3} \Re \left(\frac{1}{\tanh \left( \frac{\pi\left(u-it'(u) \right)}{\beta'} \right)} \right)+\frac{\pi }{\beta' u^2} \Re \left(\frac{1}{\sinh^2 \left( \frac{\pi\left(u-i t'(u) \right)}{\beta'} \right)} \right)   \right],
\label{induced-metric-D7}
\end{align}
where $t'(u)$ is the position of the D7-brane subject to the boundary
condition $t'(u\to \infty)= t'_f$.  We have assumed that the D7-brane
is far away from the D3-brane and neglected the quantity $\Phi$
appearing in the metric (\ref{metric-LD3-large-u}). Later we will
justify the validity of this assumption. The DBI action for the $N_f$
D7-branes then becomes
\begin{align} 
S_\text{DBI}^{\text{D7}}
&=
N_f T_7 \int d^8x e^{-(\phi-\phi_\infty)} \sqrt{ \det g_{D7}} 
= \frac{8 \pi^2 N_f  N_c \beta}{3(2\pi)^6 \lambda_5}    V_3  \int_0^\infty du \, u^4 H^{1/2} \left(1+ \left(\frac{d t'(u)}{du}\right)^2 \right)^{\frac{1}{2}}.
\label{DBI}
\end{align} 
Here $V_3$ is the volume of the $x_1,x_2,x_3$ space, and we have used
$e^{\phi_\infty}=g_s$ and $T_7=1/g_s(2\pi)^7 \alpha'^4$ which is the
tension of the D7-brane. The Euclidean action has a minimum at the
constant value $t'(u)=\beta'/2$ (mod $\beta'$).\footnote{To prove
  this, note that the minimization of the action requires (a) setting
  $dt'/du=0$ and (b) minimizing $H$. From the second line of
  \eq{induced-metric-D7} $H$, for constant $t'(u)=t'_f$ is clearly
  minimized at $t'_f=\beta'/2$ (mod $\beta'$).}
Thus the stable configuration of the D7-branes is $t'(u)= t'_f=
\beta'/2$ with $d t'(u)/du=0$. This configuration is precisely
\eq{antipodal} which corresponds to
$\theta=0$.\footnote{Ref.~\cite{Mandal:2011ws} investigated the chiral
  symmetry breaking/restoration transition by comparing the DBI action
  for the configurations of the D7-branes corresponding to these
  phases (see figure \ref{fig-D7}). In this study, the D7-branes are
  set at $t'=\beta'/2$, since this is stable for both chiral symmetry
  preserved and breaking configurations. Through the relation
  (\ref{position-theta}), it corresponds to $\theta=0$. This shows
  that the results in \cite{Mandal:2011ws} are valid for the standard
  thermal quarks.}

To calculate the $t'_f$ dependence of the DBI action, we approximate
$d t'(u)/du=0$\footnote{If we do not use the approximation $d
  t'(u)/du=0$, $c_2$ is modified but qualitative properties would not
  change.} and take $t'(u)= t'_f$. The integral \eq{DBI} evaluates as
\begin{align} 
\int_0^\infty du \, u^4 H^{1/2}=&\frac{1}{\beta^3}  \sqrt{ \frac{2 \lambda_5}{\beta}} \left(
c_0+c_2 \left( \beta t'_f-2\pi^2 \right)^2 + \cdots
 \right),
\end{align} 
where $c_0$ is a divergent constant which should be removed through a
regularization and $c_2=20.7022...$\,.  This gives us the potential of
a single D7 brane as a function of the position variable $t'_f$.  As
we found in Section \ref{impotss}, $t'_f$ is directly related to the
imaginary chemical potential $\theta$, according to the relation
(\ref{position-theta}). By substituting this relation into the above
and summing up the contributions of all D7 and $\overline{\rm
  D7}$-branes, we obtain the $\theta$ dependent classical action
\begin{align} 
 S_\text{DBI}^{{\rm D7}\ol{\rm D7}}
&= \frac{2 N_c N_f V_3}{3 \pi^2 \sqrt{ \lambda_{YM} L_4}} T^{\frac{5}{2} }  c_2 \theta^2+ \cdots  .
\label{pot-D3}
\end{align} 
Once again, we see that the choice $\theta=0$, made in
\cite{Mandal:2011ws}, and given by \eq{antipodal} 
is a consistent classical solution according to this potential.

\vspace{1ex}

\begin{figure}[H]
\begin{center}
\includegraphics[scale=1.0]{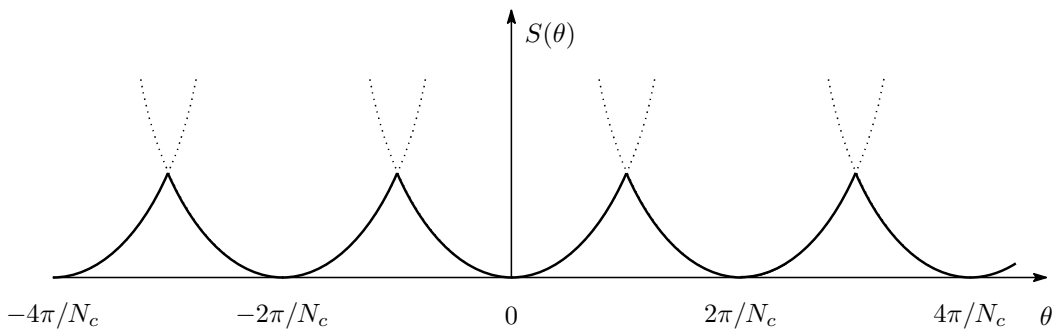}
\caption{A schematic plot of $S(\theta)$, the effective potential for the imaginary chemical potential $\theta$.
}
\label{fig-pot-theta}
\end{center}
\end{figure}

This, however, is not the end of the story.  The partition function of
QCD with an imaginary chemical potential (\ref{thermal-partition-QCD})
has a discrete symmetry
\begin{align}
Z(T;\theta) =Z(T;\theta+ 2\pi/N_c),
\label{Z_N-QCD}
\end{align}
which is a generalization of the $\mathbf{Z}_{N_c}$ symmetry of pure
Yang-Mills theory to QCD (see Appendix \ref{app-ZN-sym}). The
potential \eq{pot-D3} clearly does not respect this symmetry. To fix
this problem, we need to understand how the above $\mathbf{Z}_{N_c}$
symmetry is realized in the gravity dual. We find, following similar
considerations in \cite{Rafferty:2011hd, Aharony:1998qu,
  Witten:1998wy, Maldacena:2001ss, Yee:2009cd}, that the missing
pieces of the puzzle are provided by the dynamics of some bulk gauge
fields, namely, in our setup, the graviphoton (KK gauge field) and
$C_4$ gauge field.  We give the details in Appendix \ref{app-ZN}; the
upshot is that the dynamics of these gauge fields changes the
classical action (\ref{pot-D3}) into
\begin{align} 
S(\theta)
&= \frac{2 N_c N_f V_3}{3 \pi^2 \sqrt{\lambda_{YM} L_4}} T^{\frac{5}{2} } \left[ c_2  \min_{k \in {\bf Z}} \left(\theta - \frac{2\pi k}{N_c} \right)^2+ \cdots  \right].
\label{pot-theta-holo}
\end{align} 
The above action, plotted in figure \ref{fig-pot-theta}, obviously has
the discrete symmetry $\theta \to \theta + 2\pi/N_c$.

\noindent The potential has $N_c$ minima for $\theta$ ($0 \le \theta <
2\pi$) and indicates $N_c$ phase transitions as $\theta$ varies.  Note
that the minimum discussed above, namely $\theta=0$ (corresponding to
\eq{antipodal}) remains a minimum for the above potential as well.
Through each transition, the distance between the D7-branes and the D3-branes is shifted by $\beta'/N_c$ \footnote{To be more precise, through each transition, $\eta$ in (\ref{action-DBI+F4}) is shifted by $1/N_c$, and hence $t'_f$ is shifted by $\beta'/N_c$ due to the relation above (\ref{action-D7s}).} and they are always antipodal at the minima of the potential. 
It ensures the assumption below (\ref{induced-metric-D7}).

In \cite{Roberge:1986mm}, Roberge and Weiss obtained the following
result for the potential for an imaginary chemical potential $\theta$
in the deconfinement phase from perturbative QCD:
\begin{align} 
V_{\text{pert.}}(\theta)= \min_{k\in {\bf Z}} \left[ - \frac{1}{12} \pi^2 T^4 N_c N_f \left(1-
\frac{1}{\pi^2} \left( \theta - \frac{2\pi k}{N_c} \right)^2 \right)^2  \right].
\label{pot-theta-0}
\end{align} 
Remarkably, this potential qualitatively agrees with the holographic
result (\ref{pot-theta-holo}) we found above.

\section*{Acknowledgements}
The authors would like to thank Yoshimasa Hidaka for discussions about imaginary chemical potential. 
The work of H.~I. is supported by Iwanami Fujukai.
The work of T.~M. is supported in part by Grant-in-Aid for Scientific Research (No.\,15K17643) from JSPS.

\appendix

\section{A review of imaginary chemical potentials in field theory}
\label{app-chemical}

We summarize properties of the imaginary chemical potential which are used in the main part.

\subsection{Relation between imaginary chemical potential and periodicity of fermions}
First we discuss the connection between the imaginary chemical potential and the periodicity of the fermions.
In general, for a system described by a Hamiltonian $\wh{H}$ with fermions described by dynamical variables ${\wh\eta}_i$ (where the index $i$ refers to any set of indices of the dynamical variables including the coordinates),
we can derive the following identity 
\begin{align} \label{part2}
\tr(e^{-\bt\wh{H}+i\te\wh{N}}) & =
\int_{\vec{\eta}(\bt)=-\vec{\eta}(0)}[d\vec{\eta}{}^{\,\ast}d\vec{\eta}]
\exp\left\{
-\int_0^{\bt}\!dt\,
\left[\sum_i\eta_i{}^{\ast}\pd_{t}\eta{}_i+H(\vec{\eta}{}^{\,\ast},\vec{\eta})
-i\bt^{-1}\te\sum_i\eta_i{}^{\ast}\eta_i\right]
\right\}
\nn\\
& =
\int_{\vec{\eta}(\bt)=-e^{-i\theta}\vec{\eta}(0)}[d\vec{\eta}{}^{\,\ast}d\vec{\eta}]
\exp\left\{
-\int_0^{\bt}\!dt\,
\left[\sum_i\eta_i{}^{\ast}\pd_{t}\eta{}_i+H(\vec{\eta}{}^{\,\ast},\vec{\eta})
\right]
\right\},
\end{align}
where $t$ is the Euclidean time and $\wh{N}:=\sum_i\wh{\eta}_i{}^{\,\ast}\wh{\eta}_i$ is the fermion
number operator.  The trace on the left hand side is taken over the
fermions.  We can derive this relation in a usual manner with the
slicing in $\bt$-direction and the coherent state
representation. The nontrivial periodicity on the second line results
from the identity
$e^{i\theta\wh{N}}|\vec{\eta}\,\ran=|e^{i\theta}\vec{\eta}\,\ran$ on
the coherent state $|\vec{\eta}\,\ran$ in the trace.  Therefore we
can identify the imaginary chemical potential with the monodromy of
the fermions around the Euclidean time
circle.\footnote{\label{ftnt:monodromy}Two known special cases of
  \eq{part2} are worth noting: (a) a thermal partition with no
  imaginary chemical potential ($\theta=0$) corresponds to an
  antiperiodic boundary condition around the Euclidean time circle,
  and (b) a thermal partition function with an imaginary chemical
  potential $\theta=\pi$ corresponds to the periodic boundary
  condition. The latter observation ties up with the fact that an
  insertion of $e^{i\theta \hat N}$ in the LHS of \eq{part2} with
  $\theta=\pi$ is equivalent to inserting $(-1)^{\hat N}$, which, of
  course, converts a thermal partition function into a Witten index,
  and is hence represented by a path integral over periodic fermions.}

\subsection{Relation between imaginary chemical potential and gauge fields}\label{imgpotgauge}
Next we establish the connection between the imaginary chemical potential and the U(1) component of an external gauge field.
Let us consider a QCD-like theory with a color gauge group $G_c$ and matter fermions $\psi$ in the fundamental representation of $G_c$ and some representation of a flavor group $G_f$. 
We gauge both $G_c$ and $G_f$ with the gauge potentials $A^c$ and $A^f$ respectively, and regard $A^c$ as dynamical and $A^f$ as external.
The Lagrangian density (in the Lorentzian signature) is
\begin{align}
&{} \cL_m = -\ol{\psi}(\ga^{\mu}D_{\mu}+m)\psi \nn
\end{align}
with the following definitions:
\begin{align}
&{} \quad
\ol{\psi} := \psi^{\dagger}i\ga^0, \quad
D_{\mu}\psi := \pd_{\mu}\psi - iA^c_{\mu}\psi - iA^f_{\mu}\psi, \quad
\{\ga^{\mu}, \ga^{\nu}\} = 2\eta^{\mu\nu}, \quad
\eta_{\mu\nu} = \diag(-1,1,1,1). \nn
\end{align}
The Hamiltonian reads
\begin{align}
H_m[A^f] 
= \int\!d^3x\,[- \psi^{\dagger}A_0\psi + \ol{\psi}\ga^iD_i\psi + m\ol{\psi}\psi], \nn
\end{align}
where $A:=A^c+A^f$, and we displayed the external flavor field
$A^f$ explicitly in the Hamiltonian.
Note that the index 0 stands for the Lorentzian time while $t$ stands for the Euclidean time.

Now let us turn on only the temporal component $A_0^f$ of the external flavor gauge field $A^f$ and suppose that this $A_0^f$ is proportional to the identity matrix.
Then the first equality of \eqref{part2} with the Hamiltonian $H_m[A^f]$ and the number operator 
$\wh{N}:=\int\!d^3x\,\psi^{\dagger}\psi$
reads
\begin{align} \label{diracpart}
&\kern10pt\tr
\left(\exp\left[-\bt\wh{H}[A^f_0]+i\te \wh{N}\right]
\right)\nn\\
&{} =
\int_{\psi(\bt)=-\psi(0)} [dA^cd\psi^{\ast}d\psi] \,\, e^{-S_c[A^c]}\ \exp\left\{
-\int_0^{\bt}\!\! dt d^3x \,
[\psi^{\dagger}\pd_{t}\psi
+ \ol{\psi}(\ga^iD_i + m)\psi
- i\psi^{\dagger}A_{t}\psi 
- i\bt^{-1}\te\psi^{\dagger}\psi]
\right\} \nn\\
&= \int_{\psi(\bt)=-\psi(0)} \kern-20pt [dA^cd\psi^{\ast}d\psi] \,\, e^{-S_c[A^c]} 
\exp\left\{
-\int_0^{\bt}\!\! dt d^3x \,
[\psi^{\dagger}\pd_{t}\psi
+ \ol{\psi}(\ga^iD_i + m)\psi
- i\psi^{\dagger}\bt^{-1} \left( \beta A_{t}+ \theta \right)\psi 
]
\right\} \nn\\
&= \tr\left(\exp\left[-\bt\wh{H}[0]+i(\te + \beta A^f_t ) \wh{N}\right]
\right),
\end{align}
where we defined $A_{t}:=-iA_0$=$A^c{}_t + A^f{}_t$, 
and $S_c[A^c]$ is the Euclideanized Yang-Mills action of the color gauge field.\footnote{
The trace is taken not only over the matter fermions but also over the color gauge field.}
The final expression makes it clear that $A^f_t$ (times $\beta$)
plays the same role of an imaginary chemical potential $\theta$.

\subsection{Derivation of \eqref{chemical-gauge}}\label{derive}
We will find a string-theoretical setup which yields the thermal partition function with an imaginary chemical potential. 
The thermal partition function is given by a path integral of a Euclideanized QCD-like theory with anti-periodic fermions while a gauge theory on the D4/D8/$\ol{\rm D8}$ branes in our string-theoretical setup is a Euclidean QCD-like theory with \textit{periodic} fermions. 
If we add a diagonal gauge field $A_{t}$ on D8/$\ol{\rm D8}$, this couples with the fermions as a flavor gauge field. 
Thus the string theory side gives us the following path integral\footnote{
More precisely, a classical supergravity plus the D8/$\ol{\rm D8}$-brane actions evaluated at an on-shell configuration gives us \eqref{periodicPI} of the corresponding gauge theory, which is a QCD-like theory with periodic fermions. We perform this in the T-dual setup in Section \ref{sec-theta-dep}.}
\begin{align}\label{periodicPI}
\int_{\psi(\bt)=\psi(0)} \kern-20pt [dA^cd\psi^{\ast}d\psi] \,\, e^{-S_c[A^c]} 
\exp\left\{
-\int_0^{\bt}\!\! dt d^3x \,
[\psi^{\dagger}\pd_{t}\psi
+ \ol{\psi}(\ga^iD_i + m)\psi
- i\psi^{\dagger}\bt^{-1} \beta A_{t}\psi 
]\right\}.
\end{align}
By the second equality of \eqref{part2}, this path integral is equal to
\begin{align}
\int_{\psi(\bt)=-\psi(0)} \kern-20pt [dA^cd\psi^{\ast}d\psi] \,\, e^{-S_c[A^c]} 
\exp\left\{
-\int_0^{\bt}\!\! dt d^3x \,
[\psi^{\dagger}\pd_{t}\psi
+ \ol{\psi}(\ga^iD_i + m)\psi
- i\psi^{\dagger}\bt^{-1}(\beta A_{t} -\pi)\psi 
]\right\}.
\end{align}
This is equal to the thermal partition function with an imaginary chemical potential $\beta A_{t} -\pi$.
In other words, if we want to compute a thermal partition function with an imaginary chemical potential $\te$, then we may consider a D4/D8/$\ol{\rm D8}$-brane setup with a diagonal gauge field $\bt^{-1}(\te+\pi)$ on the D8/$\ol{\rm D8}$-branes.

\subsection{Periodicity of thermal partition function with imaginary chemical potential}
\label{app-ZN-sym}

Finally, we explain the periodicity \eqref{Z_N-QCD} of the partition function $Z(T,\te) := \tr(e^{-\bt\wh{H}+i\te\wh{N}})$, where $G_c={\rm SU}(N_c)$ and $G_f={\rm U}(N_f)$. Its path integral expression can be written down using \eqref{diracpart} as
\begin{align} \label{ztheta}
Z(T,\te) &=
\int_{\psi(\bt)=-e^{-i\te}\psi(0)}[dA^cd\psi^{\ast}d\psi] \,\,
e^{-S[\psi,\ol{\psi},A^c]-S_c[F^c]}, \\
S[\psi,\ol{\psi},A^c]
&:=
\int_0^{\bt}\!\! dt d^3x \,
[\psi^{\dagger}\pd_{t}\psi
+ \ol{\psi}(\ga^iD_i + m)\psi
- i\psi^{\dagger}A^c_{t}\psi]. \nn
\end{align}
Let us consider the following ${\rm SU}(N_c)$ transformation $g$ with a $\mathbf{Z}_{N_c}$-twisted boundary condition
\begin{align}
g(\bt,\bsx)=e^{-2\pi ik/N_c}g(0,\bsx), \quad
g(t,\bsx) \in {\rm SU}(N_c), \quad
k=0,1,\cdots,N_c-1.
\end{align}
Under this transformation, the fermions $\psi$ and the Polyakov loop $W(A^c)$ transform as
\begin{align}
\psi(t,\bsx) \mapsto \psi^g(t,\bsx)=g(t,\bsx)\psi(t,\bsx), \quad
W(A^c) \mapsto W(A^{cg}) = e^{-2\pi ik/N_c}W(A^c).
\end{align}
This transformation changes the periodicity of the matter fields as follows
\begin{align}
\psi^g(\bt,\bsx) = -e^{-i\te}\psi(0,\bsx) ~~ \longrightarrow ~~
\psi^g(\bt,\bsx)
= -e^{-i(\te+2\pi k/N_c)}\psi^g(0,\bsx).
\end{align}
Then we can derive the periodicity \eqref{Z_N-QCD} as follows:
\begin{align}
Z(T,\te) &=
\int_{\psi(\bt)=-e^{-i\te}\psi(0)}[dA^cd\psi^{\ast}d\psi] \,\,
e^{-S[\psi,\ol{\psi},A]-S_c[F^c]} \nn\\
&=
\int_{\psi^g(\bt)=-e^{-i(\te+2\pi k/N_c)}\psi^g(0)}[dA^{cg}d\psi^{\ast g}d\psi^g] \,\,
e^{-S[\psi^g,\ol{\psi}{}^g,A^{cg}]-S_c[F^{cg}]} \nn\\
&=
Z(T,\te+{2\pi k}/{N_c}),
\end{align}
where in the second equality we used the invariance of the path integral measure and the action under $g$.

\section{${\bf Z}_{N_c}$  symmetry for $\theta$ in gravity}
\label{app-ZN}

In this section, we show how the discrete ${\bf Z}_{N_c}$ symmetry
$\theta \to \theta + 2\pi/N_c$ discussed around (\ref{Z_N-QCD}) is realized in the
localized D3-brane system. The origin of this symmetry, as noted
above, is the fact that the color gauge group dual to the near
horizon geometry is $SU(N_c)$ (rather than $U(N_c)$) with a center
${\bf Z}_{N_c}$.  Related symmetries in black brane geometries have
been discussed in \cite{Aharony:1998qu, Yee:2009cd}, and our
derivation is similar to these works.

To see the ${\bf Z}_{N_c}$ symmetry in type IIB supergravity, we
consider a Kaluza-Klein decomposition on the $t'$-cycle and study the
low energy dynamics in terms of the 9 dimensional IIB supergravity.
The 9d supergravity action has the following terms\footnote{These are
  related, by T-duality along the $t'$-cycle, to the terms
\[
\frac{1}{2} \int d^{10} x \sqrt{g}\  F_{4}^2 + 
\int B_2 \wedge F_4 \wedge F_4
\]
in IIA supergravity which are employed in \cite{Yee:2009cd} to explore
the ${\bf Z}_{N_c}$ symmetry in the black D4-brane background.}
\begin{align} 
\frac{\beta'}{(2\pi)^7 \alpha'^4}\left(\frac12 
\int d^9x\ \sqrt{g}\ F_{4}^2 + \int A_1 \wedge F_4 \wedge F_4 \right),
\label{9d-action}
\end{align} 
where $F_4 = dC_3$, the three-form potential $C_3$ is the dimensional
reduction of the four-form potential $C_4$, and the one-form potential
$A_1$ is the KK gauge field from the 10 dimensional metric. We will
show that
\begin{align}
\eta \equiv \frac{1}{\beta'} \int_{u_0}^\infty A_u du 
\end{align}
takes a discrete value $ k/N_c$ for some integer $k$ at low energy---
a fact which, we will see, reflects the ${\bf Z}_{N_c}$
symmetry.\footnote{$u_0$ denotes the IR end of $u$-coordinate in the 9
  dimensional theory which is related to the position of the horizon
  of the localized D3-brane geometry.  Although the position of the
  horizon depends on $t'$ and $u$ in the original 10 dimensional
  system, within the KK-decomposition the $t'$ dependence of the
  metric is mapped to the condensation of KK non-zero modes.}

At low energies, the kinetic term of (\ref{9d-action}) reduces to
\begin{align} 
\frac{1}{2e^2}  \int_{{\bf R}^3 \times S^1_{L_4}} d^4x \,  F_{4}^2 
\end{align}
through the integration over $S^4$ and the $u$ direction.  Here $e^2$
is an effective coupling\footnote{Although $u$ integral is over an
  infinite range, the effective coupling $e^2$ is finite, as in the
  case of a similar discussion for black branes \cite{Aharony:1998qu,
    Yee:2009cd}.}.  The CS term of (\ref{9d-action}) becomes
\begin{align}
\frac{\beta'}{(2\pi)^7 \alpha'^4}\int A_1 \wedge F_4 \wedge F_4
&= \frac{1}{\beta'} \int du A_u~\frac{\beta'}{(2\pi)^4 \alpha'^2} \int_{S^4}F_4 ~ \frac{\beta'}{(2\pi)^3 \alpha'^2}  \int_{\bfR^3 \times S^1_{L_4}} F_4 
\nonumber \\
& = \mu_3 \beta' \eta N_c  \int_{\bfR^3 \times S^1_{L_4}} F_4,
\end{align} 
where we have used $ \frac{\beta'}{(2\pi)^4 \alpha'^2}\int_{S^4} F_4 =
N_c $ and $\mu_3 := (2\pi)^{-3} \alpha'^{-2} $ is the tension of a
D3-brane.  Therefore the Hamiltonian density from the action
(\ref{9d-action}) reads
\begin{align}
\mathcal{H}=\frac{e^2}{2}\left(\Pi - \mu_3 \beta' \eta  N_c \right)^2,
\end{align}
where $\Pi := F_4/e^2 + \mu_3 \eta N_c = -i \delta/\delta C_3 $ is the
conjugate momentum of $C_3$ with regard to one of the three directions
$(x_1,x_2,x_3)$ of $\bfR^3$ (see \eqref{config-IIB}) as the `time'
direction for the canonical quantization. 
Then similarly to 2d QED \cite{Coleman:1976uz, Witten:1978ka}, a basis of energy
eigenstates of this system can be taken as \cite{Aharony:1998qu, Yee:2009cd}
\begin{align}
\Psi_k = \exp \left( i k \beta' \mu_3 \int_{R^2 \times S^1_{L_4}} 
\, C_3 \right),
\end{align}
with energy eigenvalue
\begin{align}
\mathcal{H} \Psi_k= \frac{e^2\mu^2_3\beta'^2}{2}\left(k -  
\eta N_c \right)^2 \Psi_k.
\end{align}
Here $k$ must be integer-quantized. 
Note that under a shift $\eta \to \eta +1/N_c$, there is a monodromy
$\Psi_k \to \Psi_{k+1} $ accompanying a pair creation of
D3-branes\footnote{\label{ftnt:schwinger}The 
brane configuration of the created D3-branes  are given by
\begin{eqnarray}
\begin{array}{ccccccccccc}
& (t') & 1 & 2 & 3 & (4) & 5 & 6 & 7 & 8 & 9 \\
N_c~\text{D3-brane (LSD3)} &  & - & - & - & - &&&&& \\
\text{pairs of D3-branes}
& - & - & - &  & - &  &  &  & &  
\end{array}
\label{domain-wall-config}
\end{eqnarray}
where we have taken $x^3$ as the `time' direction for the canonical
quantization.  Note that this brane configuration is the T-dual of the
black D4-brane system studied in \cite{Yee:2009cd}.} (this is an
analogue of a pair creation of the electrons in 2d QED \cite{Coleman:1976uz}).  Therefore
the potential for $\eta$ becomes
\begin{align}
V(\eta)= \min_{k\in {\bf Z}} \frac{e^2\mu^2_3\beta'^2}{2}\left(k -  \eta N_c \right)^2 
\label{pot-eta}
\end{align}
and is minimized at $ \eta = k/N_c$ for some integer $k$.
Hence $\eta$ is restricted to discrete values at low energy.

Now let us show that the shift $\eta \to \eta +1/N_c$ induces the ${\bf Z}_{N_c}$ symmetry for $\theta$.
We have considered the 9 dimensional supergravity so far and ignored the KK non-zero modes.
However they should be included in the system, and importantly they always couple to the KK gauge field $A_u$ through the covariant derivative
\begin{align}
\left(\partial_u -i  \frac{ 2\pi n}{\beta'} A_u \right) \phi_n,
\end{align}
where $\phi_n$ denotes the $n$-th KK mode of field $\phi(\,=\sum_n e^{2\pi i n t'/\beta'} \phi_n)$ which symbolically denotes all fields in the nine dimensional supergravity except $A_u$.
Now $A_u$ can be absorbed into the phases of the KK modes by the field redefinition
\begin{align}
 \phi_n = \exp\left( i \frac{2\pi n}{\beta'} \int_{u_0}^{u} du' \, A_u \right) \tilde{\phi}_n. 
 \label{redefinition}
\end{align}
Using this fact, we show that we can read off the solution of supergravity with non-zero $A_u$ from the solution with  $A_u = 0$.
Since $A_u$ with no derivatives appears only at the CS term (\ref{9d-action}) after the field redefinition (\ref{redefinition}), the equation of motion for $\tilde{\phi}_n$ is the same as one in $A_u=0$ case except for the RR fields which couple to $A_u$ directly through the CS coupling.
Thus if we ignore the backreaction of these RR fields, $\tilde{\phi}_n= \left.\phi_n \right|_{A_u=0}$ is a solution of supergravity with $A_u \neq 0$, where $\left.\phi_n \right|_{A_u=0}$ is a solution of supergravity with $A_u = 0$.
Therefore in terms of the original ten dimensional coordinates 
\begin{align}
\phi(t',u) = \sum_n  \exp\left( i \frac{2\pi n t'}{\beta'} + i \frac{2\pi n}{\beta'} \int_{u_0}^{u} du' \, A_u \right) \left.\phi_n \right|_{A_u=0}.
\label{KK-decom-red}
\end{align}
is a solution of the supergravity with non-zero $A_u$.
It means that the solution is obtained just by replacing $t' \to t'+ \int^u_{u_0} du' A_u$.

Let us consider how the free energy for the imaginary chemical potential (\ref{pot-D3}) is modified by $A_u$.
The free energy depends on $\theta$ which is related to the position of the D7-branes at the boundary ($u \to \infty$) via (\ref{position-theta}).
Thus according to the above arguments, we replace $t' \to t'+  \int^\infty_{u_0} du A_u = t'+ \beta' \eta$ 
(and hence $\theta \to \theta + 2\pi \eta$ by \eqref{position-theta}) to obtain
 \begin{align} 
 S_\text{DBI}^{{\rm D7}\ol{\rm D7}}=  \frac{2 N_c N_f V_3}{3 \pi^2 \sqrt{ \lambda_{YM} L_4}} T^{\frac{5}{2} }  c_2 \left( \theta + 2\pi \eta \right)^2.
 \label{action-D7s}
\end{align}
By adding the energy (\ref{pot-eta}) for $F_4$, we obtain the $\eta$-dependent terms of the classical action
\begin{align} 
 S_\text{DBI}^{{\rm D7}\ol{\rm D7}}+S_{F_4}
&=  \frac{2 N_c N_f V_3}{3 \pi^2 \sqrt{ \lambda_{YM} L_4}} T^{\frac{5}{2} }  c_2 \left( \theta + 2\pi \eta \right)^2+\min_{k\in {\bf Z}} \frac{ L_4 V_3 e^2\mu^2_3\beta'^2}{2}\left(k -  \eta N_c \right)^2  .
\label{action-DBI+F4}
\end{align}
Then after an integral over $\eta$ the system acquires the symmetry $\theta \to \theta + 2\pi/N_c$ by compensating the shift of $\eta$ as expected in QCD (\ref{Z_N-QCD}).
By regarding this symmetry, we can rewrite the $\theta$-dependent part of the classical action (\ref{action-DBI+F4}) as\footnote{We performed an integral over $\eta$, observing that the coefficient of $\eta^2$ in the second term of \eqref{action-DBI+F4} is much larger by $N_c$ than that in the first term.} 
\begin{align} 
S(\theta)=  \frac{2 N_c N_f V_3}{3 \pi^2 \sqrt{ \lambda_{YM} L_4}} T^{\frac{5}{2} }  c_2 \min_{k\in {\bf Z}}  \left( \theta - \frac{2\pi k}{N_c} \right)^2.
\end{align}
Thus we obtain (\ref{pot-theta-holo}).

\end{document}